\documentclass[aip,jcp,reprint,twocolumn,superscriptaddress]{revtex4-2}

\pdfoutput=1

\usepackage{graphicx}
\usepackage[caption=false]{subfig}
\usepackage{amsmath}
\usepackage{amssymb}
\usepackage{color}
\usepackage[urlcolor=blue]{hyperref}

\newcommand{\am}[1]{{\textcolor{black}{#1}}}

\begin{document}


\begin{widetext}
\textit{This article may be downloaded for personal use only. Any other use requires prior permission of the author and AIP Publishing. This article appeared in J.\ Chem.\ Phys.\ \textbf{157}, 011102 (2022) and may be found at \href{https://doi.org/10.1063/5.0093598}{\color{blue}\underline{https://doi.org/10.1063/5.0093598}}.}
\end{widetext}

\vspace{1.cm}

\title{Statistics for an object actively driven by spontaneous symmetry breaking into reversible directions}

\author{Andreas M. Menzel}
\email{a.menzel@ovgu.de}
\affiliation{Institut f{\"u}r Physik, 
Otto-von-Guericke-Universität Magdeburg, Universitätsplatz 2, 39106 Magdeburg, Germany}

\date{\today}

\begin{abstract}
Propulsion of otherwise passive objects is achieved by mechanisms of active driving. We concentrate on cases in which the direction of active drive is subject to spontaneous symmetry breaking. In our case, this direction will be maintained, until a large enough impulse by an additional stochastic force reverses it. Examples may be provided by self-propelled droplets, gliding bacteria stochastically reversing their propulsion direction, or nonpolar vibrated hoppers. The magnitude of active forcing is regarded as constant, and we include the effect of inertial contributions. Interestingly, this situation can formally be mapped to stochastic motion under (dry, solid) Coulomb friction, however, with a negative friction parameter. Diffusion coefficients are calculated by formal mapping to the situation of a quantum-mechanical harmonic oscillator exposed to an additional repulsive delta-potential. Results comprise a ditched or double-peaked velocity distribution and spatial statistics showing outward propagating maxima when starting from initially concentrated arrangements. 
\end{abstract}



\maketitle

\section{Introduction}

Many of the various types of self-propelled objects that have been analyzed over the past decades feature an intrinsic polar direction setting their direction of motion. Examples are the famous biological microswimmers \textit{Escherichia coli} \cite{nishiguchi2017long, mathijssen2019oscillatory} or \textit{Chlamydomonas reinhardtii} \cite{kantsler2013ciliary}, animals like birds \cite{cavagna2010scale} or fish \cite{tunstrom2013collective, ito2021emergence}, synthetic microswimmers in the form of Janus colloidal particles \cite{jiang2010active, narinder2021active}, or vibrated polar hoppers \cite{deseigne2010collective, scholz2018inertial}. Various works have addressed the displacement statistics of such self-propelled or actively driven objects \cite{zheng2013non, sevilla2014theory, menzel2015focusing, kurzthaler2018probing, villa2020run}. When the dynamics of individual objects is described, it has become well established to represent the propulsion mechanism by an active driving force of constant magnitude \cite{ten2011brownian, pototsky2012active, ni2013pushing, speck2015dynamical, menzel2015tuned, heidenreich2016hydrodynamic}. 

However, there are self-propelled or actively driven objects not featuring any dominating, permanent polar direction that would significantly affect their direction of motion. Examples are self-propelled droplets driven by concentration gradients and/or Marangoni stresses \cite{toyota2009self, thutupalli2011swarming, yoshinaga2012drift, yabunaka2012self, schmitt2013swimming, yoshinaga2014spontaneous, seemann2016self, jin2021collective}. Spontaneous symmetry breaking of these concentration fields initiates motion. Since velocity and concentration fields are coupled, imposing an initial velocity would, vice versa, affect the direction of propulsion. 
Similarly, (roughly) isotropically shaped vibrated hoppers and nonpolar vibrated rods select their migration direction by (quasi)spontaneous symmetry breaking \cite{narayan2007long, lanoiselee2018statistical}. Using modified approaches of the Vicsek type \cite{vicsek1995novel}, the collective motion of such nonpolar objects that may feature reversal of their propulsion direction has been modeled \cite{chate2006simple, romanczuk2016emergent, menzel2016way}. Their collective behavior qualitatively differs from corresponding polar counterparts. 

Mainly, when the motion of active Brownian particles or the dynamics of microswimmers are addressed, the associated linear viscous or frictional force is assumed to dominate the dynamics. This leads to overdamped behavior. Inertial effects are neglected. Here, to also include situations where momentum effects do play a role \cite{takatori2017inertial, das2019local, lowen2020inertial, leoni2020surfing, arold2020active, caprini2021inertial, sprenger2021time}, we explicitly keep the inertial terms. 
Additionally, the objects are subject to a stochastic force. It represents, for instance, thermal fluctuations in the active Brownian case or variations involved in the sensitive bouncing sequences of hoppers on vibrating plates. 

We mainly assume the motion to be confined to one dimension along a line (or a circle of sufficiently large radius). Corresponding setups could be realized for 
self-propelled droplets on appropriately prepared surfaces or boundaries, similarly as for self-propelled polar Janus particles \cite{das2015boundaries, simmchen2016topographical} or previous considerations on passive objects \cite{wang2009anomalous, wang2012brownian}. For hoppers, corresponding tracks have been realized using confining walls \cite{blair2003vortices, volfson2004anisotropy}. \am{Example trajectories for two-dimensional systems are presented as well.}

It turns out that, under all these conditions, the resulting equations of motion formally agree with those of stochastic motion under (dry, solid) Coulomb friction \cite{degennes2005brownian, goohpattader2010diffusive, touchette2010brownian, menzel2011effect, touchette2012exact}. The magnitude of self-propulsion maps to the strength of Coulomb friction, however, with the central difference of a negative 
sign, which leads to qualitatively different results. Specifically, the velocity distribution shows a central v-shaped ditch and thus becomes double-peaked or features fully separated maxima at elevated strengths of active driving. The displacement distribution develops two oppositely (outward) moving maxima. We calculate effective diffusion coefficients and confirm them by explicit agent-based simulations. 

Next, in Sec.~\ref{sec_stochastic-equations}, we introduce the stochastic equations of motion in a Langevin picture and present the corresponding Fokker-Planck equation. In Sec.~\ref{sec_velocity-distr}, the stationary velocity distribution is derived. Moreover, the associated velocity-dependent part of the Fokker-Planck equation is analyzed by formal mapping to the problem of a quantum-mechanical harmonic oscillator supplemented by a repulsive $\delta$-potential. From there, we calculate the resulting diffusion coefficients. 
We investigate the spatial displacement statistics in Sec.~\ref{sec_spatial} by direct numerical iteration in time of the Fokker-Planck equation. At elevated magnitude of active driving, it leads to the described outward propagation of maxima. Conclusions are drawn in Sec.~\ref{conclusion}.

\section{Stochastic equations of motion}
\label{sec_stochastic-equations}

The active propulsion mechanism of the considered 
object 
constantly drives it into the present direction of motion. This scenario is described by 
Langevin-type equations of motion, 
\begin{eqnarray}
m\,\frac{\mathrm{d}v}{\mathrm{d}t} &=&
{}-\zeta\,v +A\,\sigma(v) +\Gamma(t),
\label{eq_Lang_v}
\\[.1cm]
\frac{\mathrm{d}x}{\mathrm{d}t} &=& v.
\label{eq_Lang_x}
\end{eqnarray}
Here, $x$ denotes the position of the object along its one-dimensional path, $v$ its velocity, $m$ its mass, and $t$ represents time. As motivated above, we maintain explicitly the inertial contributions and do not confine ourselves to overdamped motion. $\zeta$ parameterizes the coefficient of (viscous) linear friction. The next term $A\,\sigma(v)$ represents the active driving mechanism of magnitude $A>0$. Here, $\sigma(v)$ is the sign-function so that $\sigma(v)=1$ if $v>0$, $\sigma(v)=0$ if $v=0$, and $\sigma(v)=-1$ if $v<0$. 
Finally, $\Gamma(t)$ includes the stochastic force acting at time $t$ on the object. We consider a $\delta$-correlated form of Gaussian distribution 
as a frequently employed approximation. 
Thus, $\langle\Gamma(t)\rangle=0$ and $\langle\Gamma(t)\Gamma(t')\rangle=2K\,\delta(t-t')$, where $K$ sets the strength of the stochastic force. In the case of objects of colloidal size, thermal fluctuations of the environment determine $K$. One then frequently uses the magnitude set by the fluctuation--dissipation theorem in the passive situation of $A=0$, that is $K=\zeta\,k_B T$, where $k_B$ represents the Boltzmann constant and $T$ temperature. For vibrated hoppers, $K$ is set by the amplitude of surface vibrations. 

We note that 
$A\,\sigma(v)$ in Eq.~(\ref{eq_Lang_v}) is of the same functional form as the frictional contribution associated with (dry, solid) Coulomb friction, frequently written as $-\Delta\,\sigma(v)$. Consequences of the latter during stochastic motion have been evaluated to quite some extent \cite{degennes2005brownian, hayakawa2005langevin, goohpattader2010diffusive, mettu2010stochastic, touchette2010brownian, menzel2011effect, touchette2012exact}. 
Our case of active motion therefore corresponds to a situation of Coulomb friction of negative friction coefficient $\Delta\equiv-A<0$. The opposite sign has essential consequences, which are the subject of this work. 

Besides the Langevin-type equations, we 
address the associated Fokker-Planck equation
\begin{equation}\label{eq_fp}
\partial_t f = 
\left\{ {}-v\partial_x 
+ \partial_v \left[v-A\,\sigma(v)\right] 
+ \partial_v^2
\right\} \!f,
\end{equation}
where $f=f(x,v,t)$ denotes the probability distribution to find the object at time $t$ at a certain position $x$ with a certain velocity $v$. This distribution is normalized so that $\int_{-\infty}^{\infty}\mathrm{d}x\int_{-\infty}^{\infty}\mathrm{d}v \,f(x,v,t)=1$. To obtain the equation in the presented form, we have rescaled $x$ by $(Km)^{1/2}\zeta^{-3/2}$, $v$ by $(K/m\zeta)^{1/2}$, $t$ by $m/\zeta$, $A$ by $(K\zeta/m)^{1/2}$, and $f$ by $\zeta^2/K$, maintaining normalization. Thus the strength of active driving $A$ is our only remaining parameter. It sets the deviation from a passive system.

\section{Velocity distribution and diffusion coefficient}
\label{sec_velocity-distr}

%
%
%
%
%
%

We begin by addressing the pure velocity distribution $f_v(v,t)=\int_{-\infty}^{\infty}\mathrm{d}x\,f(x,v,t)$. The corresponding dynamic equation results from Eq.~(\ref{eq_fp}) by integration over the whole range of the spatial variable $x$. 

\begin{figure}
\centerline{\includegraphics[width=.9\columnwidth]{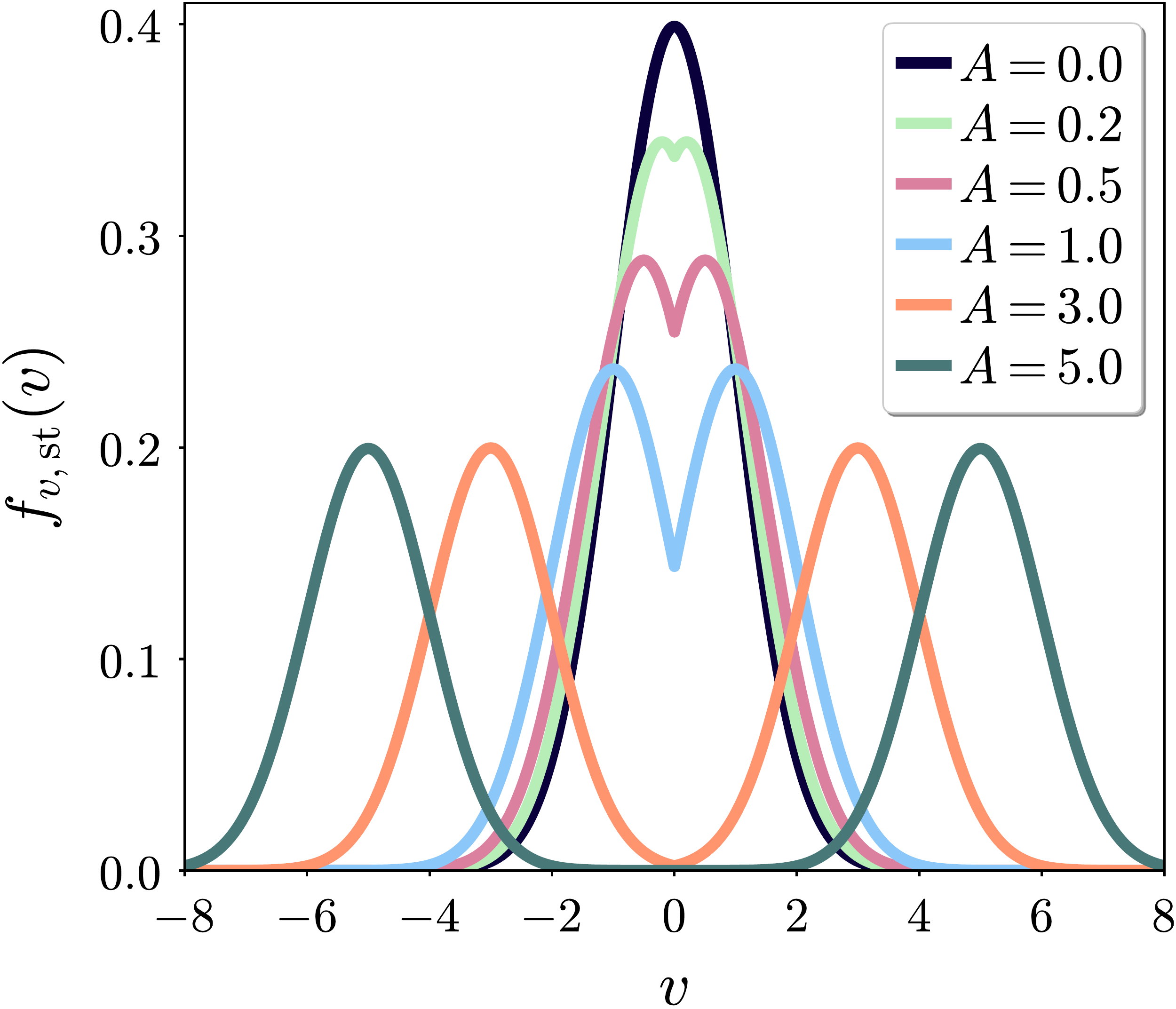}}
\caption{
Stationary velocity distribution $f_{v,\mathrm{st}}(v)$, see Eq.~(\ref{eq_fvst}), for different strengths of active driving $A$. For $A=0$, we obtain the regular Gaussian velocity distribution of passive Brownian motion. With increasing $A>0$, a v-shaped ditch emerges, before a rather bimodal velocity distribution results for $A\gtrsim3$. 
}
\label{fig_fvst}
\end{figure}

First, we note that a stationary velocity distribution $f_{v,\mathrm{st}}(v)$ exists, which satisfies Eq.~(\ref{eq_fp}) for $\partial_tf_{v,\mathrm{st}}(v)=0$, 
\begin{equation}\label{eq_fvst}
f_{v,\mathrm{st}}(v) =
\frac{\mathrm{e}^{-\frac{v^2}{2}+A\,|v|}}
{\sqrt{2\pi}\,\mathrm{e}^{\frac{A^2}{2}}
\left[1+\mathrm{erf}\left(\frac{A}{\sqrt{2}}\right)\right]}. 
\end{equation}
This distribution develops a v-shaped ditch at $v=0$ with increasing $A>0$ as depicted in Fig.~\ref{fig_fvst}. 
Two maxima develop in $f_{v,\mathrm{st}}(v)$, and the distribution becomes increasingly bimodal.
\am{For $A=3$, $f_{v,\mathrm{st}}(v)$ at $v=0$ decays to approximately one percent of its maximum.}
Thus, for $A\gtrsim3$ the two peaks are basically separated from each other. 
It becomes more and more unlikely that the propulsion direction of the object gets reversed. Long experimental waiting times are then necessary to observe ergodicity. 
Active driving is simply too strong to let the stochastic force oftentimes reverse the propulsion direction. 

Multiplying Eq.~(\ref{eq_fp}) by $[f_{v,\mathrm{st}}]^{-1/2}$ from the left and defining $\bar{f}_v=[f_{v,\mathrm{st}}]^{-1/2}f_v$, we obtain
\begin{equation}\label{eq_fp_fbar}
\partial_t \bar{f}_v = 
\left\{ 
\frac{1}{2}\left[ 1-2\,A\,\delta(v) \right]
-\frac{1}{4}\left[ v-A\,\sigma(v) \right]^2
+ \partial_v^2
\right\} \!\bar{f}_v. 
\end{equation}
$\delta(v)$ denotes the Dirac $\delta$-function, and we have used that $\partial_v\sigma(v)=2\,\delta(v)$. The equation for $\bar{f}=[f_{v,\mathrm{st}}]^{-1/2}f$ has the same form with an extra term $-v\,\partial_x\bar{f}$ on the right-hand side.

In fact, the operator on the right-hand side of Eq.~(\ref{eq_fp_fbar}) is Hermitian. We may seek to determine its eigenvalues $-\mu$ and associated eigenfunctions $\psi_{\mu}(v)$. Then, the solution to Eq.~(\ref{eq_fp_fbar}) reads
\begin{equation}
\bar{f}_v(v,t) = \sum_{\mu} a_{\mu}\,\psi_{\mu}(v)\,\mathrm{e}^{-\mu\,t},
\end{equation}
where $a_{\mu}$ are corresponding expansion coefficients. 

For $v\gtrless0$, we define a shift in variables $\tilde{v}=v\mp A$. Thus, the eigenvalue problem that we need to solve becomes
\begin{equation}\label{eq_psitild}
{}-\partial_{\tilde{v}}^2\tilde{\psi}_{\mu}(\tilde{v})+\left[\frac{1}{4}\tilde{v}^2+A\,\delta(\tilde{v}\pm A)\right]\tilde{\psi}_{\mu}(\tilde{v}) =
\left(\mu+\frac{1}{2}\right)\tilde{\psi}_{\mu}(\tilde{v})
\end{equation}
for corresponding eigenfunctions $\tilde{\psi}_{\mu}(\tilde{v})$. We note that this equation is of identical form as the Schr\"odinger equation in quantum mechanics, here for a harmonic potential plus a repulsive $\delta$-potential at $\tilde{v}=\mp A$. A similar relation was found for the case of Coulomb friction, where, however, the $\delta$-potential is pinning \cite{menzel2011effect}. 

The general solutions to Eq.~(\ref{eq_psitild}) are given by the parabolic cylindrical functions $D_{\mu}(\tilde{v})$. At $v=0$ an additional condition for the derivative of $\psi_{\mu}(v)$ arises, as generally for a $\delta$-contribution to the potential in the Schr\"odinger equation. 
It implies 
\begin{equation}
\frac{D_{\mu+1}(A)}{D_{\mu}(A)} 
+ \frac{D_{\mu+1}(-A)}{D_{\mu}(-A)} = -A, 
\end{equation}
which identifies the eigenvalues $\mu$. This relation has been identified before for a quantum-mechanical harmonic oscillator exposed to an additional $\delta$-potential \cite{janke1988statistical}. An alternative expression can be derived using the relation $D_{\mu}'(-A)=\mu \,D_{\mu-1}(-A)+A\,D_{\mu}(-A)/2$ that leads to 
\begin{equation}\label{eq_EWeven}
\mu=0 \quad \mathrm{or} \quad D_{\mu-1}(-A)=0. 
\end{equation}
It identifies $\mu=0$ as the lowest eigenvalue and needs to be solved numerically for the remaining eigenvalues $\mu$. 
\am{In contrast to the passive case of $A=0$, providing the analogy to a pure quantum-mechanical harmonical oscillator, the eigenvalues $\mu$ associated with $A\neq0$ are generally not of integer value.}

\am{Together,} requiring continuity at $v=0$, we \am{construct} the associated normalized \am{even} eigenfunctions \am{as}
\begin{equation}\label{eq_EFeven}
\psi_{\mu}(v)=
C_{\mu}\,D_{\mu}(|v|-A), 
\end{equation}
where
\begin{equation}
C_{\mu} =
 \left[\int_{-\infty}^{\infty}D_{\mu}(|v|-A)^2\,\mathrm{d}v
 \right]^{-\frac{1}{2}} .
\label{eq_EFevenC}
\end{equation}
%
\am{Additional eigenvalues $\mu$ follow when the eigenfunctions vanish for $v=0$ so that they are not affected by the $\delta$-potential. Keeping Eq.~(\ref{eq_EFeven}) for $v>0$ as an ansatz, this implies for $v\rightarrow0$ that}
\begin{equation}\label{eq_EWodd}
D_{\mu}(-A)=0. 
\end{equation}
\am{To construct the associated odd eigenfunctions, we extend the ansatz to $v<0$ as}
\begin{equation}\label{eq_EFodd}
\psi_{\mu}(v)=
C_{\mu}\,\sigma(v)\,D_{\mu}(|v|-A), 
\end{equation}
where
\begin{equation}\label{eq_EFoddC}
C_{\mu} =
{ \left[\int_{-\infty}^{\infty}\sigma(v)\,D_{\mu}(|v|-A)^2\,\mathrm{d}v\right]^{-\frac{1}{2}}
}.
\end{equation}
Moreover, we note from Eqs.~(\ref{eq_EWeven}) and (\ref{eq_EWodd}) that the eigenvalues associated with the corresponding even and odd eigenfunctions emerge in pairs that differ by integer $1$ (except for the eigenvalue $\mu=0$). This relation has already been noted in the case of Coulomb friction \cite{touchette2010brownian}. 

\begin{figure}
\centerline{\includegraphics[width=.9\columnwidth]{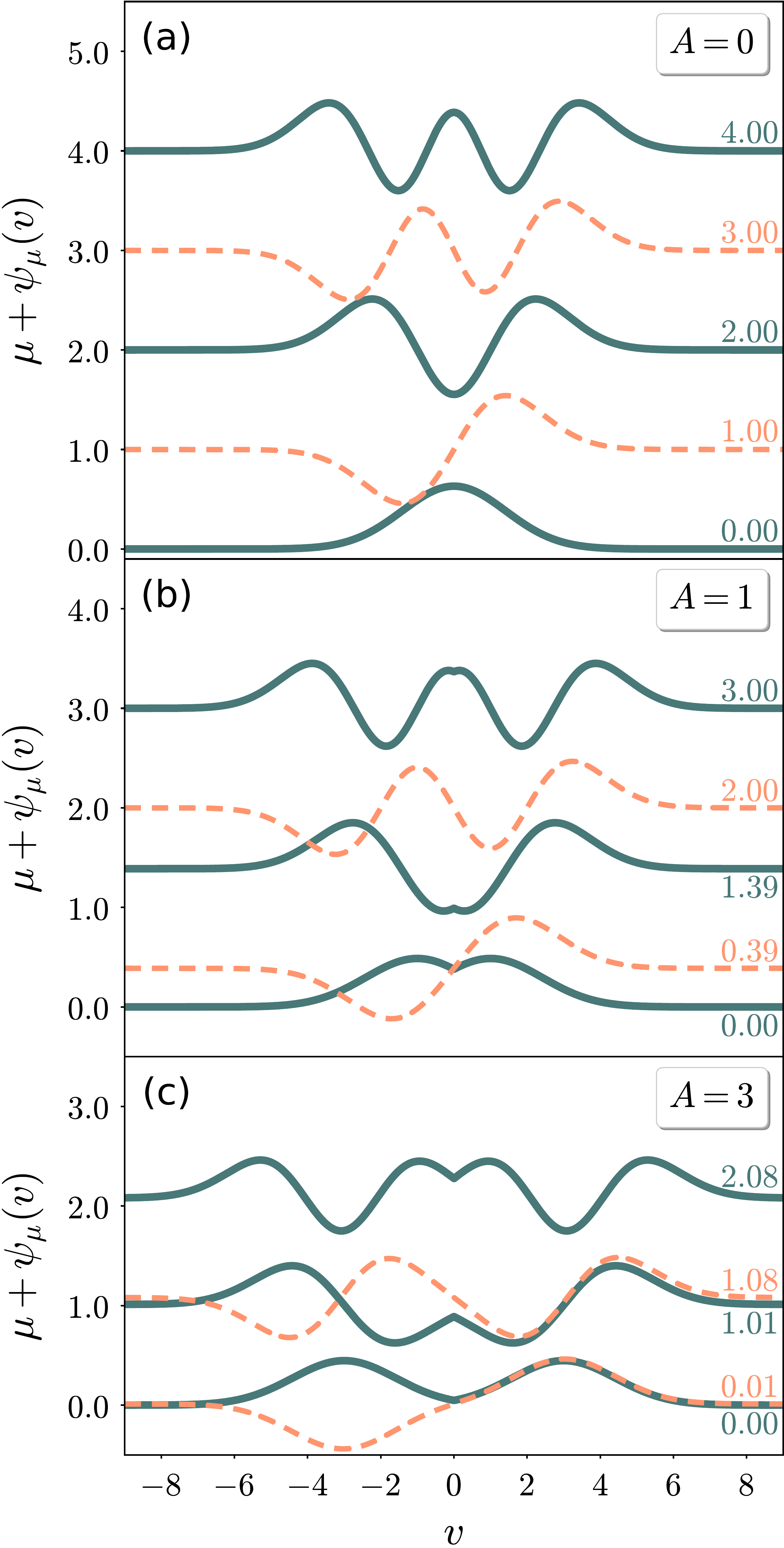}}
\caption{
Eigenfunctions $\psi_{\mu}(v)$ according to Eqs.~(\ref{eq_EWeven})--(\ref{eq_EFevenC}) and (\ref{eq_EWodd})--(\ref{eq_EFoddC}), associated with the lowest five eigenvalues $\mu$ for strengths of active driving (a) $A=0$, (b) $A=1$, and (c) $A=3$. The eigenfunctions are shifted along the ordinate according to $\mu$\am{, which is indicated on the right-hand side of each curve}. Odd eigenfunctions are marked by dashed lines. For even eigenfunctions, kinks due to active driving $A\neq0$ appear at $v=0$. 
}
\label{fig_EF}
\end{figure}

Using Mathematica \cite{mathematica2020}, we have determined numerically from Eqs.~(\ref{eq_EWeven}) and (\ref{eq_EWodd}) eigenvalues up to $\mu\lesssim50$ for $A=0$, $A=0.1$, $A=0.3$, $A=0.5$, $A=0.7$, $A=1$, and $A=3$. Some associated eigenfunctions obtained via Eqs.~(\ref{eq_EWeven})--(\ref{eq_EFevenC}) and (\ref{eq_EWodd})--(\ref{eq_EFoddC}) are depicted in Fig.~\ref{fig_EF}. 
For the eigenfunctions resulting from Eq.~(\ref{eq_EFeven}) for $A>0$, the influence by the $\delta$-potential becomes obvious from the kink at $v=0$. As expected, the odd eigenfunctions determined from Eq.~(\ref{eq_EFodd}) pass the origin smoothly. 
Moreover, we have checked orthonormality by numerical integration on a test basis. 

From the spectrum of eigenvalues and associated eigenfunctions, we can calculate the diffusion coefficient $D$ under active driving using the velocity autocorrelation function $\left\langle v(0)v(t)\right\rangle$. To this end, the propagator of the Fokker-Planck equation for $f_v(v,t)$ is expressed in eigenfunctions $\psi_{\mu}(v)$ \cite{degennes2005brownian}. We find
\begin{eqnarray}
D&=&
\int_0^{\am{\infty}} \mathrm{d}s\,\left\langle v(0)v(s)\right\rangle 
\nonumber
\\[.1cm]
&=&
\sum^{\mathrm{(odd)}}_{\mu} \frac{1}{\mu}
\left[ \int_{-\infty}^{\infty}\mathrm{d}v\,
\sqrt{f_{v,\mathrm{st}}(v)}\,v\,\psi_{\mu}(v)
\right]^2.
\label{eq_D}
\end{eqnarray}
Here, since $f_{v,\mathrm{st}}(v)$ is an even function with respect to $v=0$ and $v$ is odd, only the odd eigenfunctions $\psi_{\mu}(v)$ resulting from Eqs.~(\ref{eq_EWodd})--(\ref{eq_EFoddC}) contribute under the integral, as 
the remark ``(odd)'' indicates on the sum symbol. 

\begin{figure}
\centerline{\includegraphics[width=.9\columnwidth]{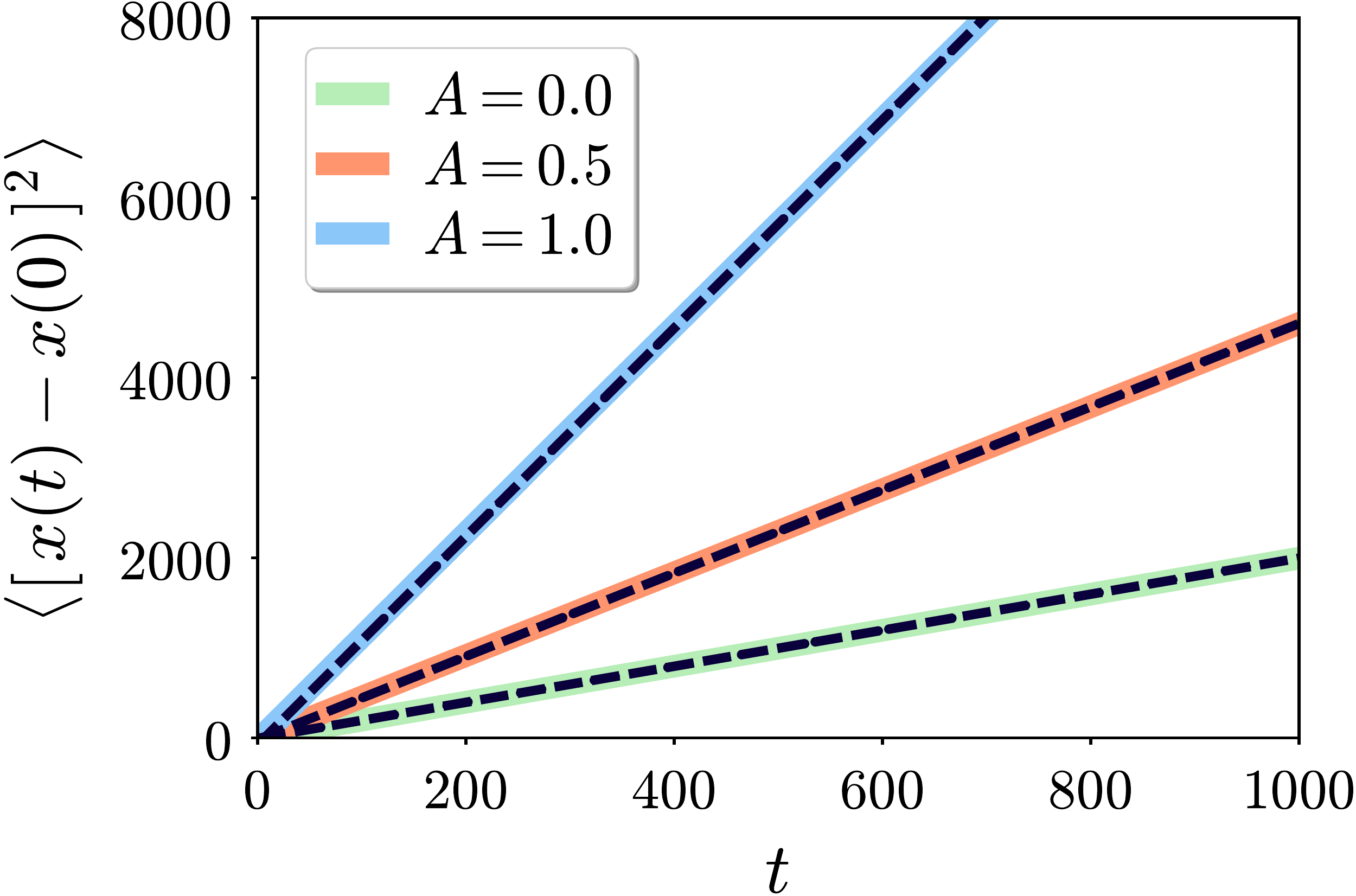}}
\caption{
Mean-squared displacement (msd) obtained from agent-based simulations according to Eqs.~(\ref{eq_Lang_v}) and (\ref{eq_Lang_x}) as a function of time $t$. Results were averaged over $10^6$ trajectories for different magnitudes of active driving $A$. The diffusion coefficients $D_{\mathrm{msd}}$ were obtained from linear fits (dashed lines) of the expression $\langle[x(t)-x(0)]^2\rangle=2\,D_{\mathrm{msd}}\,t$ to the curves in the range $800\leq t\leq1000$. They match well the coefficients $D$ calculated explicitly via Eq.~(\ref{eq_D}), see Tab.~\ref{tab_D}. [Time step in the agent-based simulations $\mathrm{d}t=0.001$.]
}
\label{fig_D}
\end{figure}

Including eigenvalues of $\mu\lesssim50$, we calculate from Eq.~(\ref{eq_D}) for various strengths of active driving $A$ the diffusion coefficients $D$, see Tab.~\ref{tab_D}. 
\begin{table}
\setlength{\tabcolsep}{5pt}
\begin{tabular}{|l||c|c|c|c|c|c|c|}
\hline
$A$  &  0.0  &  0.1  &  0.3  &  0.5  &  0.7  &  1.0  &  3.0
\\ \hline\hline
$D$  &  1.00 & 1.17 & 1.64 & 2.31 & 3.29 & 5.76 & 792.65 
\\ \hline\hline
$D_{\textrm{msd}}$  &  1.00 & 1.18 & 1.64 & 2.31 & 3.30 & 5.78 & 797.13  
\\ \hline
\end{tabular}
\caption{For different strengths of active driving $A$, effective diffusion coefficients $D$ are calculated from Eq.~(\ref{eq_D}). They are compared to corresponding coefficients $D_{\mathrm{msd}}$ obtained from fits to the mean-squared displacements obtained from direct agent-based simulations of Eqs.~(\ref{eq_Lang_v}) and (\ref{eq_Lang_x}). All relative deviations are less than one percent.}
\label{tab_D}
\end{table}
%
%
To verify the results, 
we performed explicit agent-based simulations of Eqs.~(\ref{eq_Lang_v}) and (\ref{eq_Lang_x}). The associated temporal evolution of the mean-squared displacement is depicted in Fig.~\ref{fig_D} for a few cases. \pagebreak
Corresponding diffusion coefficients $D_{\mathrm{msd}}$ are extracted from the curves by linear fits $\langle[x(t)-x(0)]^2\rangle=2\,D_{\mathrm{msd}}\,t$, which are listed in Tab.~\ref{tab_D} for comparison. They match well the values obtained from Eq.~(\ref{eq_D}) with relative deviations of less than one percent.

\section{Spatial distribution}
\label{sec_spatial}

To study the time evolution of the spatial distribution $f_x(x,t)$, we solve Eq.~(\ref{eq_fp}) numerically for $f(x,v,t)$. 
We use finite differences and employ a second-order upwind scheme to address convective contributions. As an initial condition, we multiply the stationary velocity distribution in Eq.~(\ref{eq_fvst}) by a narrow spatial Gaussian distribution of standard deviation $0.1$. 
At selected times $t$ of evaluation, we calculate $f_x(x,t)=\int_{-\infty}^{\infty}\mathrm{d}v\,f(x,v,t)$. 

\begin{figure}
\centerline{\includegraphics[width=.9\columnwidth]{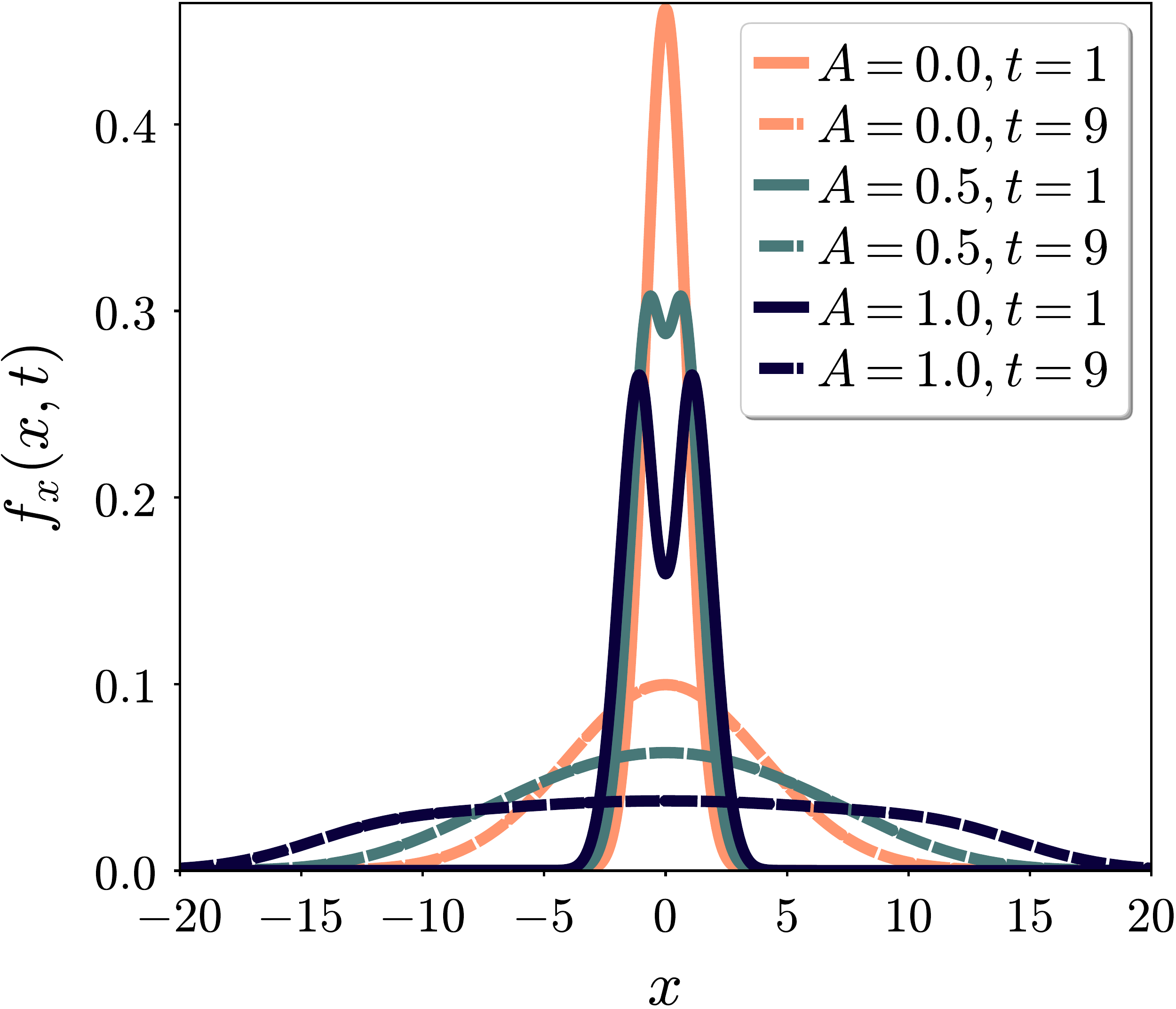}}
\caption{
Temporal evolution of the spatial distribution $f_x(x,t)$ under active driving $A=0.5$ and $A=1$ when compared to regular passive diffusion for $A=0$. The quicker spreading of the distributions with increased active driving is obvious, in agreement with the elevated magnitudes of the diffusion coefficients in Tab.~\ref{tab_D}. 
[Time step in the simulations: $\mathrm{d}t=10^{-5}$, number of velocity bins $N_v=1.500$, number of spatial bins $N_x=10.000$, velocity increments $\mathrm{d}v=0.01$, and spatial increments $\mathrm{d}x=0.01$.]
}
\label{fig_active-passive-diffusion}
\end{figure}

\begin{figure}
\centerline{\includegraphics[width=.9\columnwidth]{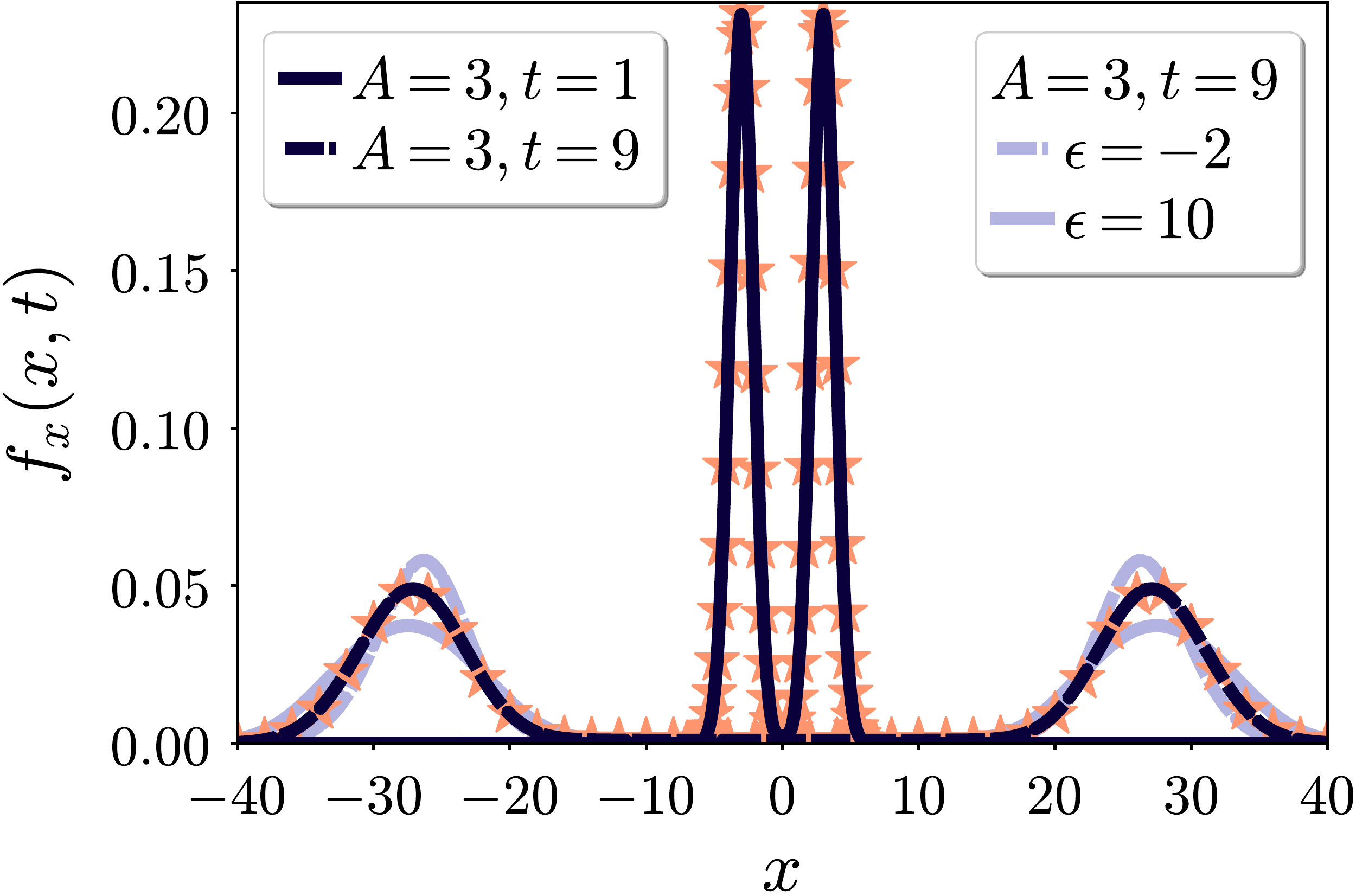}}
\caption{
Temporal evolution of the spatial distribution $f_x(x,t)$ for elevated magnitude of active driving $A=3$. On the considered time scales, pronounced maxima appear in the distribution that propagate outward with a speed of approximately $A$. Their magnitude decays over time. 
[Time step in the simulations: $\mathrm{d}t=10^{-5}$, number of velocity bins $N_v=1.500$, number of spatial bins $N_x=10.000$, velocity increments $\mathrm{d}v=0.01$, and spatial increments $\mathrm{d}x=0.01$.] 
Results obtained from corresponding agent-based simulations are indicated by stars and are in good agreement. [Number of agents $N=10^7$ and time step $\mathrm{d}t=0.001$.] 
\am{The light solid curve marks the result starting from the same initial conditions under mutual repulsive interactions $\epsilon\,\delta(x-x')$ between actively driven objects, here for $\epsilon=10$. 
Effects of mutual attractive interactions are represented by the light dashed curve, here for $\epsilon=-2$.}
}
\label{fig_prop}
\end{figure}

As expected by the elevated diffusion coefficients, see Tab.~\ref{tab_D}, the spatial distribution spreads significantly quicker as a function of time under active driving $A>0$ than for regular passive diffusion, see Fig~\ref{fig_active-passive-diffusion}. 
Moreover, with increasing magnitude of active driving, we observe outward propagating fronts and associated outward propagating density peaks, 
see Fig.~\ref{fig_prop}. 
When active driving becomes strong compared to diffusion, 
Eq.~(\ref{eq_Lang_v}) suggests propagation of speed $|v|\approx A$. Indeed, this is roughly the speed of the maxima in Fig.~\ref{fig_prop} that originate from a concentration around $x=0$ at $t=0$. Still, diffusive processes are at work and the propagating maxima decay in magnitude over time. 
\am{Mutual interactions between actively driven objects may support or hinder this decay. 
In the Fokker-Planck approach, we address the influence of a very basic general interaction potential in rescaled units, namely $V(x-x')=\epsilon\,\delta(x-x')$ for one object at position $x$ and one at $x'\,$ \cite{menzel2015focusing}. Using the mean-field approximation, it leads to an additional contribution $\epsilon\,\partial[f(x,v,t)\int\mathrm{d}v'\,\partial f(x,v',t)/\partial x]/\partial v$ in Eq.~(\ref{eq_fp}). $\epsilon>0$ expresses mutual (steric) repulsion, if two objects are located at the same position. 
$\epsilon<0$ marks mutual attraction. While the former accelerates the decay of the peaks, the latter slows it down, see Fig.~\ref{fig_prop}.}

\begin{figure}
\centerline{\includegraphics[width=.9\columnwidth]{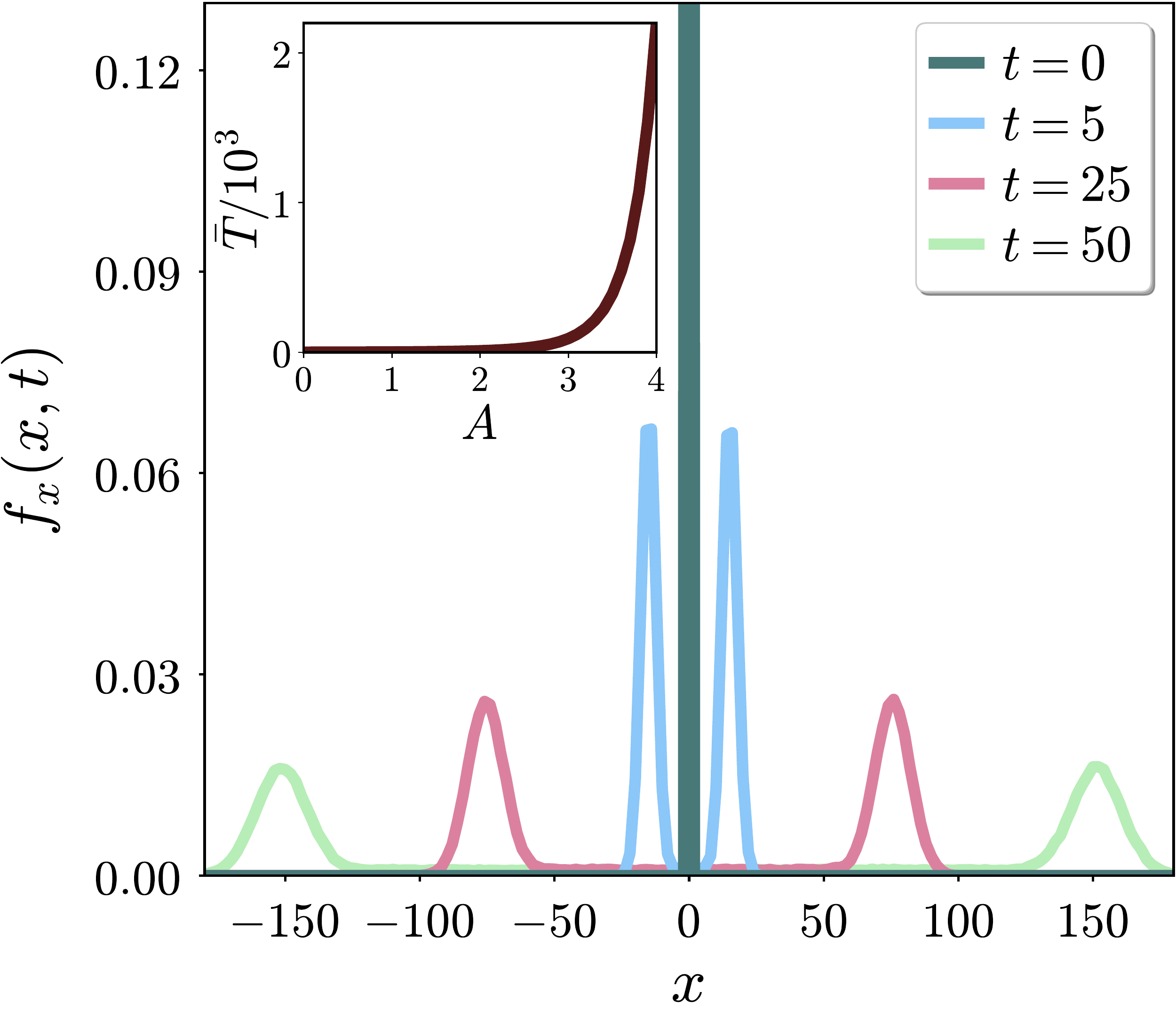}}
\caption{
Displacement statistics obtained via agent-based simulations for an elevated magnitude of active driving $A=3$ at longer times. Outward propagating maxima in the spatial statistics are observed.  
[Number of agents in the agent-based simulations $N=10^7$ and time step $\mathrm{d}t=0.001$.] \am{The inset shows that the mean time $\bar{T}$ to reverse the velocity direction strongly increases for $A\gtrsim3$. 
[$N=10^5$, $\mathrm{d}t=0.001$.]}
}
\label{fig_longtime}
\end{figure}

\begin{figure}
\centerline{\includegraphics[width=.9\columnwidth]{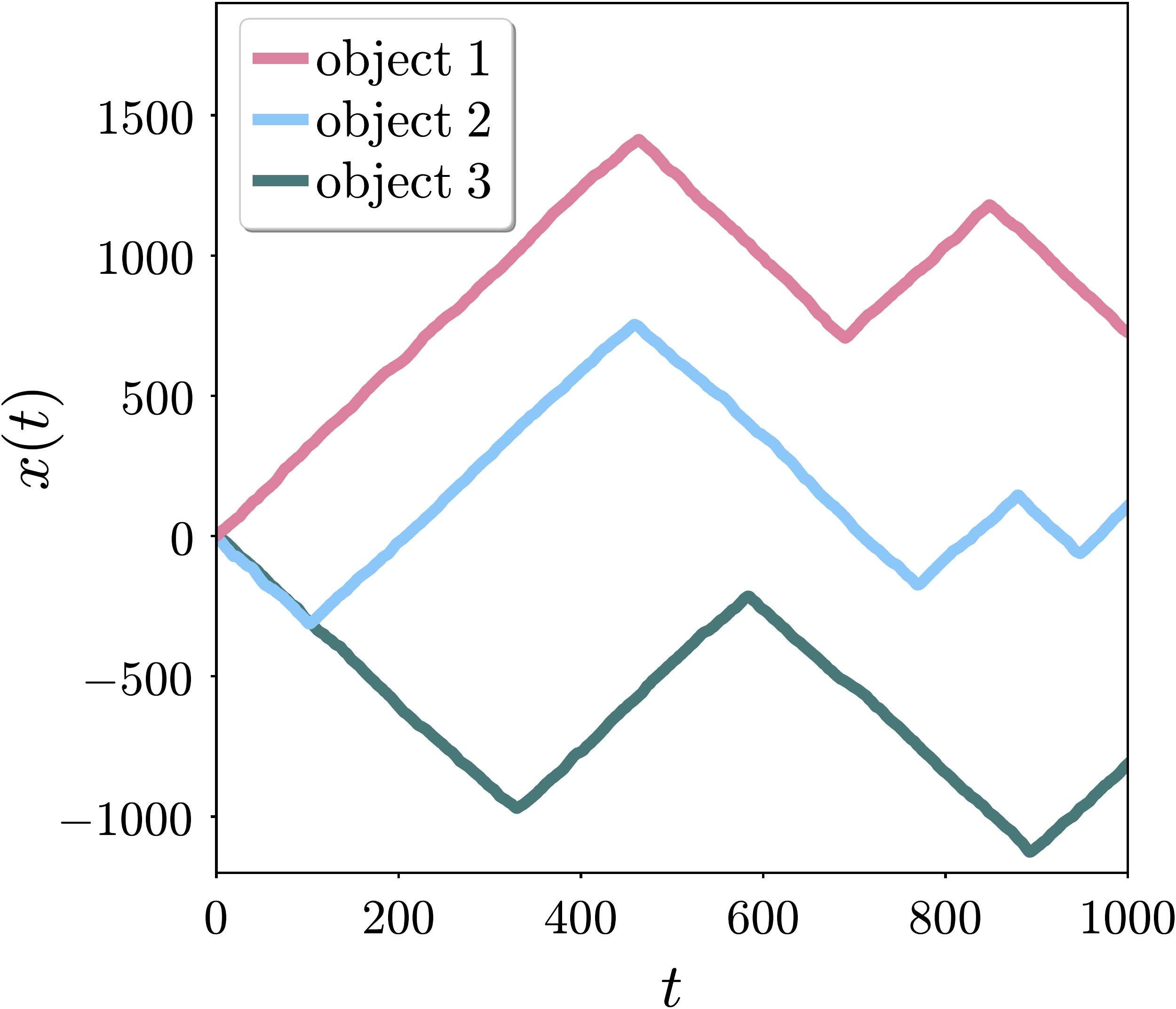}}
\caption{
Three example trajectories obtained by numerical integration of Eqs.~(\ref{eq_Lang_v}) and (\ref{eq_Lang_x}) are depicted for three objects at an elevated magnitude of active driving $A=3$. The curves illustrate motion of speed $|v|\approx A$ and rather rare events of reversing the propagation direction. 
[Time step in the agent-based simulations $\mathrm{d}t=0.001$.]
}
\label{fig_example-trajectories}
\end{figure}

Results from agent-based simulations of Eqs.~(\ref{eq_Lang_v}) and (\ref{eq_Lang_x}) for individual objects match well those obtained from the Fokker-Planck equation Eq.~(\ref{eq_fp}), see Fig.~\ref{fig_prop} for a comparison and Fig.~\ref{fig_longtime} for longer simulation times. To obtain proper statistics, elevated numbers of objects need to be considered. 
The agent-based picture provides an illustrative explanation of why the propagating fronts appear. With rising magnitude of active driving $A$ in Eq.~(\ref{eq_Lang_v}), it becomes increasingly difficult for the stochastic force $\Gamma(t)$ of given average strength to reverse the propagation direction. Thus, an individual object will in fact propagate relatively persistently in one direction with speed $|v|\approx A$, before at some point the stochastic force manages to reverse the propagation direction. \am{The mean time that it takes to achieve such a reversal is depicted in the inset of Fig.~\ref{fig_longtime}. To calculate it, we initialized agent-based simulations with velocities according to $f_{v,\mathrm{st}}(v)$, see Eq.~(\ref{eq_fvst}). In line with Fig.~\ref{fig_fvst}, where $f_{v,\mathrm{st}}(v)$ basically drops to zero at $v=0$, this mean time to reverse the direction strongly increases for $A\gtrsim3$.} Corresponding example trajectories for $A=3$ are depicted in Fig.~\ref{fig_example-trajectories}. 

\am{This feature persists when we turn to corresponding two-dimensional trajectories. 
We consider the actively driven motion of an object that features an intrinsic nonpolar axis $\mathbf{\hat{n}}$. If there is no active driving perpendicular to $\mathbf{\hat{n}}$, and if the coefficient of linear friction $\zeta_{\perp}$ for motion perpendicular to $\mathbf{\hat{n}}$ is sufficiently large or if genuine dry (solid) friction of large enough magnitude prevails in this perpendicular direction ($A_{\perp}<0$), then motion of the object is approximately confined along the nonpolar axis $\mathbf{\hat{n}}$. 
The resulting direction along $\mathbf{\hat{n}}$ is selected by spontaneous symmetry breaking and may vary over time. While the velocity along $\mathbf{\hat{n}}$ is described by Eq.~(\ref{eq_Lang_v}), the spatial position is updated according to 
\begin{equation}\label{eq_vecr}
\frac{\mathrm{d}\mathbf{r}}{\mathrm{d}t} = v\,\mathbf{\hat{n}}. 
\end{equation}}

\am{In two dimensions, we parameterize $\mathbf{\hat{n}}=(\cos\varphi, \sin\varphi)$. The dynamics of $\varphi$ is set by the angular velocity $\omega$, 
\begin{equation}
\frac{\mathrm{d}\varphi}{\mathrm{d}t} = \omega, 
\end{equation}
while the dynamics of $\omega$ is given by
\begin{equation}\label{eq_omega}
J\,\frac{\mathrm{d}\omega}{\mathrm{d}t} = 
{}-\zeta_{\mathrm{r}}\,\omega + \Gamma_{\mathrm{r}}(t). 
\end{equation}
Here, $J$ is the moment of inertia, $\zeta_{\mathrm{r}}$ the coefficient of linear rotational friction, and $\Gamma_{\mathrm{r}}(t)$ a stochastic rotational force of Gaussian distribution satisfying $\langle\Gamma_{\mathrm{r}}(t)\rangle=0$ and $\langle\Gamma_{\mathrm{r}}(t)\Gamma_{\mathrm{r}}(t')\rangle=2\,K_{\mathrm{r}}\,\delta(t-t')$.}

\am{We rescale all quantities as listed at the end of Sec.~\ref{sec_stochastic-equations}. Moreover, we rescale $J$ by $(m^3K_{\mathrm{r}}/\zeta^3)^{1/2}$ and $\zeta_{\mathrm{r}}$ by $(m\,K_{\mathrm{r}}/\zeta)^{1/2}$. Consequently, $K_{\mathrm{r}}$ is removed from Eq.~(\ref{eq_omega}).}

\begin{figure}
\centerline{\includegraphics[width=.9\columnwidth]{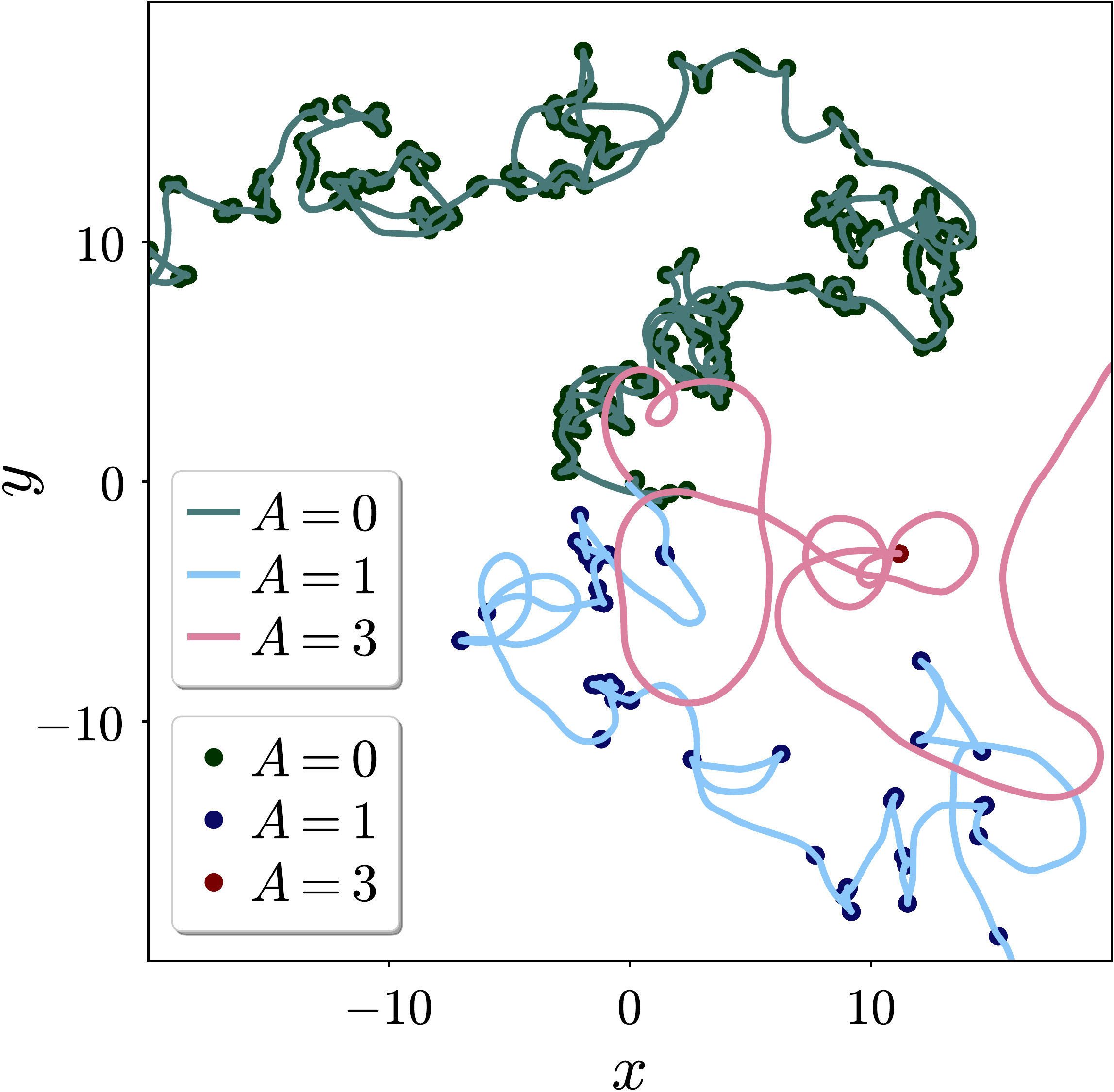}}
\caption{
\am{Three example trajectories on a two-dimensional plane obtained by numerical integration of Eqs.~(\ref{eq_Lang_v}) and (\ref{eq_vecr})--(\ref{eq_omega}) are shown for three different magnitudes of active driving $A$. Dots and cusps on the trajectories mark events of velocity reversal, which become rare with increasing $A$. 
[$J=1$, $\zeta_{\mathrm{r}}=1$, and time step in the agent-based simulations $\mathrm{d}t=0.001$ in rescaled units.]}
}
\label{fig_2D}
\end{figure}

\am{Resulting example trajectories are depicted in Fig.~\ref{fig_2D}. Cusps on these trajectories clearly indicate events of reversing the propulsion direction. Such cusps become rare with increasing magnitude of $A$, in agreement with Figs.~\ref{fig_fvst}, \ref{fig_longtime}, and \ref{fig_example-trajectories}.}
\am{Experiments with appropriately prepared vibrated rods featuring a polar or nonpolar axis \cite{mohammadi2020dynamics} could distinguish between corresponding trajectories.}

\section{Conclusions}
\label{conclusion}

We have considered the velocity and spatial displacement statistics of objects subject to active driving, its direction being determined by spontaneous symmetry breaking. 
Stochastic contributions may affect the driving direction.
We do not exclude the role of inertial terms, in agreement with recent considerations on actively driven hoppers and so-called microflyers \cite{lowen2020inertial, sprenger2021time}. As a result, we find double-peaked stationary velocity statistics with a v-shaped central ditch. For elevated active driving, the velocity distribution becomes rather bimodal. In addition, we then observe in the spatial distribution two maxima propagating outward with approximately the speed determined by active driving, starting from a concentrated central density peak. 

To proceed, we evaluated the corresponding Fokker-Planck equation. If we confine ourselves to the velocity statistics, the situation can be mapped formally to the problem of a quantum-mechanical harmonic oscillator supplemented by a repulsive $\delta$-potential. This analogy allowed us to calculate from the associated eigenfunctions the resulting diffusion coefficients, which we verified by explicit agent-based simulations. The displacement statistics were evaluated by direct numerical iteration of the Fokker-Planck equation and agent-based simulations. 

Interestingly, our situation can be mapped to the scenario of a stochastically driven object under viscous damping and additional (dry, solid) friction of the Coulomb type \cite{degennes2005brownian, goohpattader2010diffusive, touchette2010brownian, menzel2011effect, touchette2012exact}. Yet, in our case, the Coulomb friction parameter is of negative sign, which leads to the different phenomenology as summarized above. We remark that a negative friction-associated parameter \am{was introduced into} the study of active objects in different \am{frameworks before. Previous works on active Brownian \cite{schweitzer1998complex, dunkel2001thermodynamics, lindner2008critical, romanczuk2012active} and self-propelled deformable \cite{ohta2009deformable, hiraiwa2011dynamics, menzel2012soft} particles considered a negative linear (viscous) friction parameter, which in our case is positive. Thus, active driving is linear to the current speed in these situations, in contrast to our active driving of constant magnitude \cite{ten2011brownian, pototsky2012active, ni2013pushing, speck2015dynamical, menzel2015tuned, heidenreich2016hydrodynamic}. While in our case linear friction plays the major role in damping, frictional forces cubic in the velocity are frequently employed for damping in models of negative linear friction parameter \cite{erdmann2000brownian, lindner2008critical, romanczuk2012active}. Another, related but} more complex context \am{concerns} actively driven mesoscale turbulence \cite{wensink2012meso, dunkel2013minimal, reinken2018derivation}. There, it is the effective viscosity of the active suspension that turned to a negative value. In a sense, our interpretation of \am{constant} active driving as \am{dry (solid)} friction of inverted sign extends such concepts to a different situation.  

Our analytical considerations have been restricted to one-dimensional motion. Corresponding confinement can be realized in experiments. It is not straightforward to generalize for higher dimensions the formal mapping to the described quantum-mechanical problem. \am{Nevertheless, evaluations of the theory are reasonable in two dimensions, for instance for vibrated hoppers, if these objects feature a nonpolar axis of active driving with possible stochastic reversal of their velocity direction, and in two or three dimensions, for example for single- or multi-flagellated swimming bacteria that may stochastically reverse their propulsion direction \cite{magariyama2005difference, theves2013bacterial}.} 
\am{Elongated bacteria that glide on a substrate along their body axis by spontaneous symmetry breaking represent another two-dimensional example \cite{wada2013bidirectional}.}

\begin{acknowledgments}
\vspace{-.4cm}
The author thanks the Deutsche Forschungsgemeinschaft (German Research Foundation, DFG) for support through the Heisenberg Grant No.~ME 3571/4-1. 
\end{acknowledgments}
\vspace{-.3cm}

\section*{Data availability}
\vspace{-.4cm}
The data that support the findings of this study are available within the article and/or result from solving 
the equations described in the text.
\vspace{-.3cm}

\section*{Author declaration}
\vspace{-.4cm}
The author has no conflicts to disclose. 
\vspace{-.3cm}

\section*{References}

\vspace{-.7cm}


%

\end{document}